Annette Bussmann-Holder* and Hugo Keller

# High-temperature superconductors: underlying physics and applications

**Abstract:** Superconductivity was discovered in 1911 by Kamerlingh Onnes and Holst in mercury at the temperature of liquid helium (4.2 K). It took almost 50 years until in 1957 a microscopic theory of superconductivity, the so-called BCS theory, was developed. Since the discovery a number of superconducting materials were found with transition temperatures up to 23 K. A breakthrough in the field happened in 1986 when Bednorz and Müller discovered a new class of superconductors, the so-called cuprate high-temperature superconductors with transition temperatures as high as 135 K. This surprising discovery initiated new efforts with respect to fundamental physics, material science, and technological applications. In this brief review the basic physics of the conventional low-temperature superconductors as well as of the high-temperature superconductors are presented with a brief introduction to applications exemplified from high-power to low-power electronic devices. Finally, a short outlook and future challenges are presented, finished with possible imaginations for applications of room-temperature superconductivity.

**Keywords:** applications; history of superconductivity; superconductivity.

## 1 General overview

The concept of superconductivity was introduced by Heike Kamerlingh Onnes in 1911 in order to characterize a material with zero electrical resistance, unknown at that time [1]. The discovery of superconductivity was unintended, and the main reason to conduct the experiment was the liquefaction of the last noble gas helium. In order to utilize liquid helium the resistance of an as pure as possible metal was measured, namely mercury, with the aim to explore the temperature dependence of the resistivity at very low temperature. At that time there existed several theories and speculations about the low temperature behavior of metals, which ranged from continuously decreasing to unusual upturns, requiring experimental verification. Unexpectedly, liquid mercury showed zero resistance below 4.2 K which corresponds to –268°C. Kamerlingh Onnes was awarded with the Nobel Prize for Physics in 1913, however, not for the discovery of superconductivity, but for his investigations of material properties at very low temperature.

## 2 Discovery and interpretation of low temperature superconductivity: the BCS theory

After demonstrating superconductivity in mercury, a series of metals were measured and found to be superconducting as well, yet only at ultralow temperatures. The highest transition temperature $T_c$ in elemental metals was seen in niobium with a $T_c$ of 9.25 K. In order to enable applications of superconductors, liquid helium has to be used which is rather expensive. Only in the 1950[th] the A15 superconductors have been discovered, reaching a maximum $T_c$ of 23.2 K [2]. In 1986 Bednorz and Müller [3] found a new class of superconductors, the ceramic so-called cuprate high-temperature superconductors, with the at present highest $T_c$ of 135 K at ambient pressure. They were awarded with the Nobel Prize for Physics only 1 year later, the fastest recognition ever. In Table 1 an incomplete list of various types of superconductors together with their values of $T_c$ is shown.

Before going into more details of the phenomenon of superconductivity, some fundamental explanations are needed, related to metals, insulators and their properties. These are shown in Fig. 1, where the decisive difference between the two material classes is shown. In a metal, free charges, namely electrons, are present which provide the electrical current. In contrast, in an insulator all charges are strongly bound to the ion and cannot move

*Corresponding author: Annette Bussmann-Holder, Max-Planck Institute for Solid State Research, Heisenbergstrasse 1, D-70569 Stuttgart, Germany, e-mail: a.bussmann-holder@fkf.mpg.de
Hugo Keller: Physik-Institut der Universität Zürich, Winterthurerstrasse 190, CH-8057 Zürich, Switzerland

Table 1: Incomplete list of various types of superconductors (SC) and their superconducting transition temperature $T_c$.

| Type of SC | Substance | $T_c$ (K) |
|---|---|---|
| Simple metal SC | Al | 1.17 |
| | In | 3.41 |
| | Hg | 4.20 |
| | Sn | 3.72 |
| | Pb | 7.20 |
| | Nb | 9.25 |
| A15 SC | $Nb_3Sn$ | 18.0 |
| | $Nb_3Ge$ | 23.2 |
| Fullerene SC | $C_{60}Rb_3$ | 31 |
| Cuprate SC | $La_{2-x}Sr_xCuO_4$ | 38 |
| | $YBa_2Cu_3O_7$ | 93 |
| | $Bi_2Sr_2Ca_2Cu_3O_{10}$ | 107 |
| | $Tl_2Ba_2Ca_2Cu_3O_{10}$ | 125 |
| | $HgBa_2Ca_2Cu_3O_8$ | 135 |
| Magnesium diboride SC | $MgB_2$ | 39 |
| Iron-based SC | FeSe | 8 |
| | $LaO_{0.89}F_{0.11}FeAs$ | 26 |
| | $Sr_{0.5}Sm_{0.5}FeAsF$ | 56 |

Table 2: List of some typical metals (M), insulators (I), and superconductors (SC).

| Substance | M/I/SC | $\rho$ (ohm m) | $T_c$ (K) |
|---|---|---|---|
| Silver | M | $1.6 \times 10^{-8}$ | – |
| Copper | M | $1.7 \times 10^{-8}$ | – |
| Gold | M | $2.4 \times 10^{-8}$ | – |
| Aluminium | M/SC | $2.8 \times 10^{-8}$ | 1.17 |
| Lead | M/SC | $2.2 \times 10^{-7}$ | 7.20 |
| Mercury | M/SC | $9.8 \times 10^{-7}$ | 4.20 |
| Porcelain | I | $5 \times 10^{12}$ | – |
| Rubber | I | $6 \times 10^{14}$ | – |
| Quartz (fused) | I | $7.5 \times 10^{17}$ | – |

The resistivity $\rho$ at room temperature and the superconducting transition temperature $T_c$ are also given.

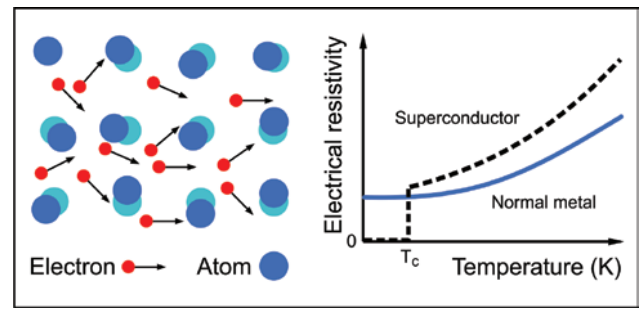

Fig. 2: (Left panel) Scattering of the electrons by collisions with the ions in the medium. (Right panel) Temperature dependence of the resistivity of a typical metal (solid line) and a superconductor (dashed line).

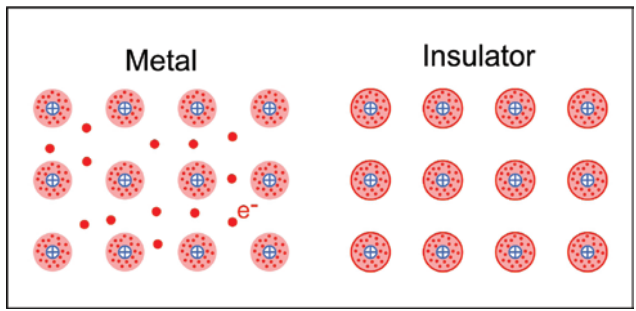

Fig. 1: Schematic representation of a metal (left) and an insulator (right). In the metal the carriers (electrons, small red circles) move almost freely in the matrix of the positive ions, whereas in the insulator they are tightly bound to the ionic core.

freely through the crystal. If an electric field is applied to a conductor, the charges move and provide the current. In Table 2 typical metals, insulators and superconductors are listed. The electrical conductance of a material can be characterized by its resistivity $\rho$ which has small values for good conductors (metals) and very large values for bad conductors (insulators) as indicated in Table 2. It is interesting to note, that the metals with the highest electrical conductivity, namely silver, copper, and gold are not superconductors, whereas metals with a low electrical conductivity such as aluminium, lead, and mercury are superconducting. How can one visualize the resistivity which is typically observed in any metal? This is illustrated in Fig. 2 (left). In a metal the electrons which carry the current are scattered by the atoms/ions which vibrate due to thermal motion. Since this motion is reduced with decreasing temperature, the resistance is reduced as well and levels off in the low temperature regime (Fig. 2 (right)). It remains, however, finite at all temperatures. This is in strong contrast to a superconductor, where the resistivity of the superconductor becomes unmeasurably small at the transition temperature $T_c$ as shown by the dashed line in Fig. 2 (right). The well-known experimental result of the discovery of superconductivity in mercury from Heike Kamerlingh Onnes and Gilles Holst is shown in Fig. 3 [1]. The distinct difference between a normal metal and a superconductor is exemplified in Fig. 2 (right). While in a metal the electrical resistivity decreases with temperature and reaches an almost temperature independent finite value at low temperature, in a superconductor the resistivity is typically larger than in a metal at high temperature – signaling the behavior of a metal with low electrical conductivity – and at $T_c$ it drops to zero signifying perfect electrical conductance. Two fundamental properties are intimately interrelated with the superconducting state, namely (i) perfect conductance and (ii) the so-called Meissner-Ochsenfeld effect by which a magnetic field is completely expelled from the superconductor, whereas in the normal state it penetrates the superconductor (see Fig. 4). The Meissner-Ochsenfeld effect is the reason for magnetic levitation.

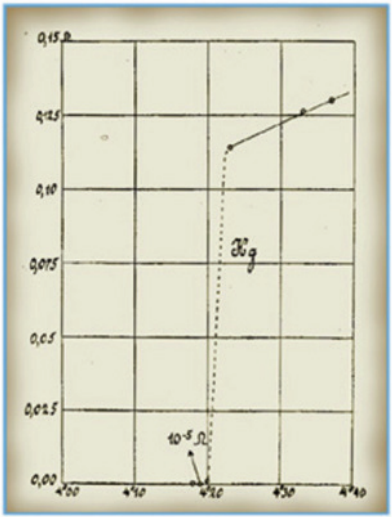

**Fig. 3:** The original experimental result of the resistivity of mercury from Heike Kamerlingh Onnes and Gilles Holst [1].

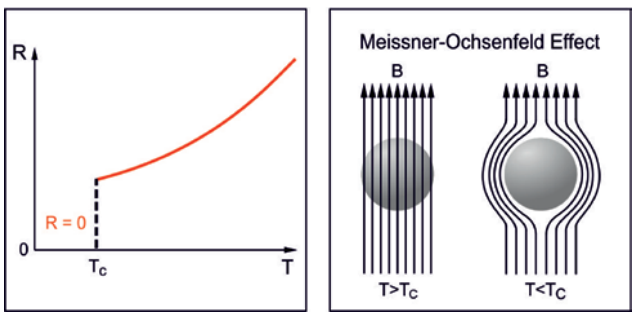

**Fig. 4:** Fundamental properties of a superconductor. (left panel) (i) Zero resistance (perfect conductor). For $T < T_c$ a superconductor is a perfect conductor with no resistance. (right panel) (ii) Meissner-Ochsenfeld effect (ideal diamagnet). For temperatures $T > T_c$ the magnetic field $B$ penetrates the superconductor, whereas for $T < T_c$ it is expelled from the sample.

It is important to emphasize that the complete expulsion of the magnetic field represents only an ideal state, whereas in reality the magnetic field penetrates over a small shell of the sample surface and decays exponentially on a material typical length scale, the magnetic penetration depth.

The understanding of the amazing phenomenon of superconductivity took long, namely 45 years. It is intimately connected with the observation of an isotope effect on $T_c$ which was reported independently by two groups in 1950 [4, 5]. In an isotope experiment the mass of one of the constituent lattice ions is exchanged by an isotope, e.g. natural abundant mercury with average atomic mass of 200.6 ($^{200.6}$Hg) is replaced by almost isotope pure $^{198}$Hg [4]. For the sample with the heavier isotope mass 200.6 the transition temperature $T_c$ is lower with the shift in $T_c$ being proportional to the inverse square root of the ionic mass [4, 5]. Almost simultaneously with the publication of the experimental finding of the isotope effect Fröhlich proposed that superconductivity is intimately related to the interaction between electrons and ions in form of the electron-phonon interaction [6]. However, from the above we have already related the electrical resistivity of metals to the scattering of electrons from the ions, which is in apparent contradiction to Fröhlich's proposal. The missing link was provided by Cooper who postulated that the charge carriers in the superconducting state are not made up of single electrons, but of electron pairs (Cooper pairs) with opposite spin and momentum [7]. The glue for binding these pairs was then provided by the lattice in terms of the electron-phonon interaction. The complete theoretical formulation of this theory has been summarized by Bardeen, Cooper and Schrieffer in 1957 and named BCS theory after them [8]. They were awarded with the Nobel Prize for Physics in 1972. In this context it is important to mention that in no other field in physics so many Nobel prizes were awarded as in the field of superconductivity.

In order to visualize the attractive pairing interaction in superconductors one has to imagine an electron being surrounded by positively charged ions. As a consequence of the attractive interaction between the electron and the neighboring ions, the lattice is deformed around the electron (Fig. 5). In this way the negative charge of the electron is screened and appears as an effective positively charged cloud, thus attracting a second electron. The paired state is a macroscopic coherent quantum state with unique properties. The distance between the paired electrons, i.e. their coherence length, is very large in conventional superconductors and can easily exceed 100 nm (typical lattice constants are of the order of 0.5–1 nm). This implies that in these superconductors many pairs coexist in a unit volume and consequently also overlap (Fig. 6 (left)). In addition, for this state to be favored over the normal state, its energy must be lower and the energy gain stemming from the pair formation being called the energy gap. This can be measured directly and provides an excellent tool to confirm superconductivity.

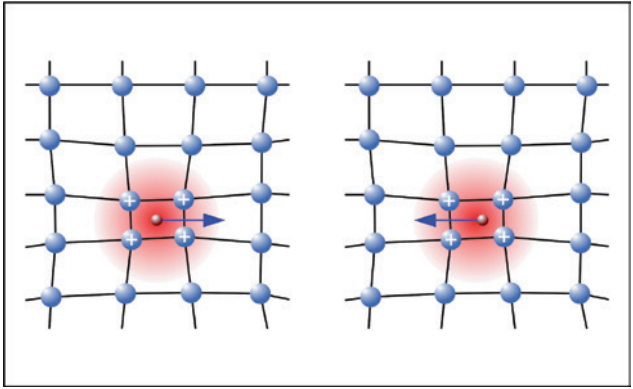

**Fig. 5:** Schematic representation of the attractive interaction between two electrons due to the electron-lattice interaction forming a Cooper pair. The first electron polarizes the lattice (left). The second electron is attracted to the concentration of positive charge left behind the first electron (right).

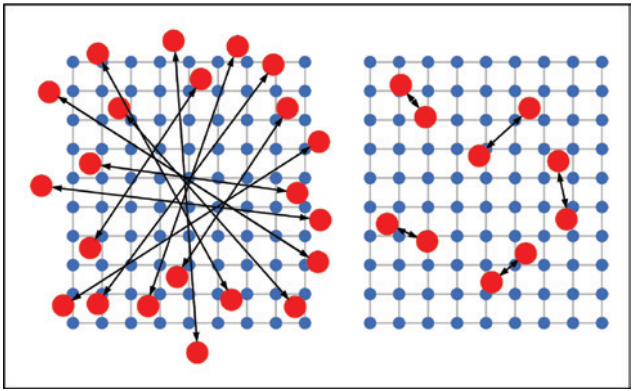

**Fig. 6:** (Left) Overlapping electron pairs in a conventional low-temperature superconductor. (Right) Electron pairs in a cuprate high-temperature superconductor.

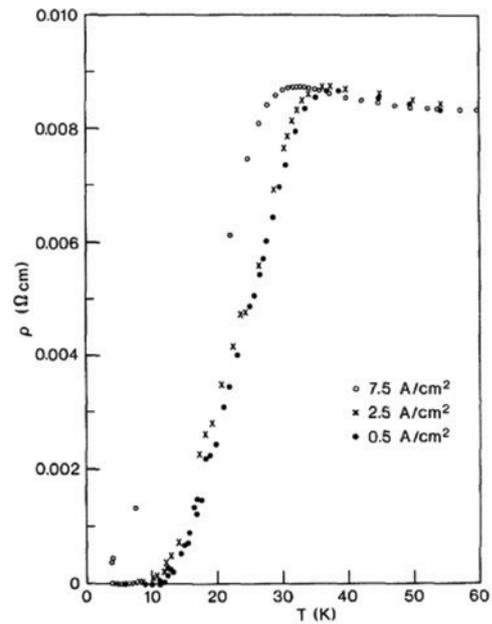

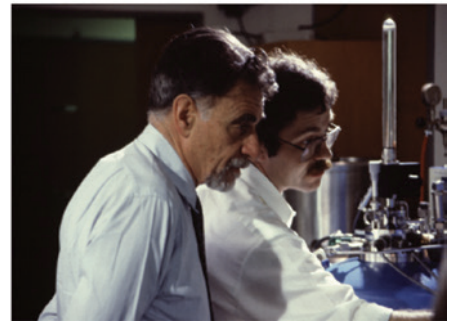

**Fig. 7:** Discovery of superconductivity in the $La_{2-x}Ba_xCuO_4$ system by Bednorz and Müller in 1986. (Top) Low-temperature resistivity of a $La_{2-x}Ba_xCuO_4$ sample recorded at different current densities (taken from [3]). The drastic drop of the resistivity at $T_c \approx 30$ K signals the appearance of superconductivity. (Bottom) J. G. Bednorz (right) and K. A. Müller (left) at the time of their discovery.

In these conventional superconductors the values of $T_c$ never exceeded 25 K thereby limiting applications drastically. This remained true for 75 years, since Heike Kamerlingh Onnes discovery and a true breakthrough was achieved in 1986 when Bednorz and Müller discoverd high-temperature superconductivity in a ceramic material, namely the cuprates (see Fig. 7) [3].

## 3 Discovery of several families of high-temperature superconductors

With the discovery of ceramic cuprates a new era in the research of superconductivity set in, since within a few months only $T_c$ reached record temperatures exceeding easily liquid nitrogen values and supporting further applications of superconductors. However, a drawback of cuprates is their ceramic nature which makes the material brittle, easily breakable and difficult to handle. Meanwhile these intricacies have partially been overcome and possible applications and realizations are given below.

While for many years it was assumed that cuprates are unique and the only true high temperature superconductors, in 2001 Akimitsu and coworkers [9] discovered another high-temperature superconductor, namely $MgB_2$ with a $T_c$ of 39 K. Even though this value lies far below the highest values of cuprates, it exceeds enormously those of conventional ones. In 2006 another breakthrough was achieved by Hosono and coworkers [10] who demonstrated high-temperature superconductivity in iron based layered compounds. The maximum $T_c$ is 56 K, still well below that of cuprates. More recently, in 2015, Eremets and coworkers [11] managed to turn $H_2S$ into a superconductor by applying extremely high pressures and reaching a $T_c > 200$ K. This finding by itself is extremely interesting, since it shows that hydrogen bonded systems are potential candidates to realize room-temperature superconductivity. However, the pressures needed to achieve superconductivity are beyond limits where applications are realizable. In common to the above systems is the fact that they are layered materials. The cuprates and iron based superconductors share a similar phase diagram where the number of added carriers determines the magnitude of $T_c$. In addition, both are magnetic as long as they are insulating. The development of $T_c$ for various conventional and high-temperature superconductors as function of the year of their discovery is depicted in Fig. 8, and an incomplete list of various types of superconductors is shown in Table 1.

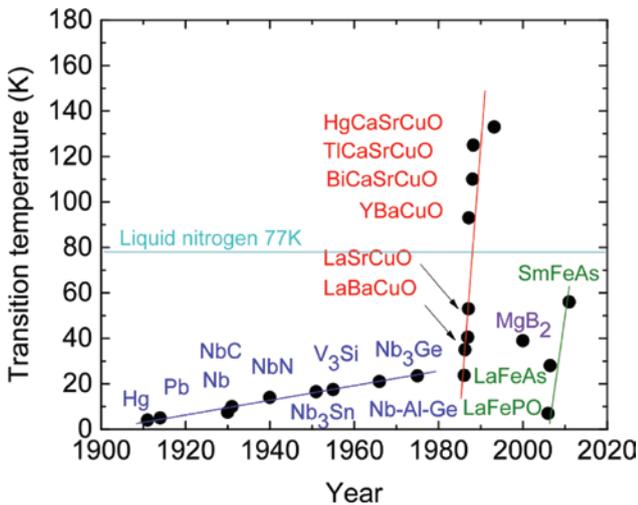

**Fig. 8:** The development of $T_c$ with time for conventional and cuprate superconductors.

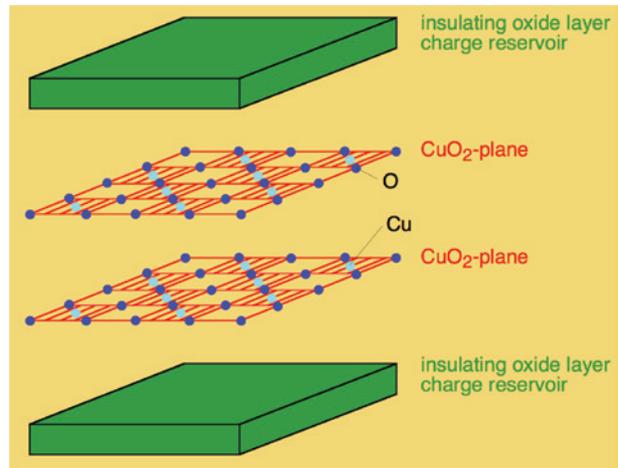

**Fig. 9:** Schematic structure of a cuprate HTS superconductor.

## 4 Physics of high temperature superconductors

Several families of cuprate HTS superconductors have been discovered. They all have a layered structure consisting of $CuO_2$ layers separated by insulating layers (see Fig. 9). Superconductivity occurs in the $CuO_2$ layers, and the insulating layers serve as charge reservoirs supplying the charge carriers to the $CuO_2$ layers.

The discovery of cuprate superconductors was not accidental as advertised by many scientists, but born from the idea that a very strong electron-lattice interaction is needed in order to arrive at high values of $T_c$. One would argue that the BCS theory already addresses this issue, however, limitations in this theory are that the lattice frequency, the number of free carriers and the coupling between carriers and the lattice are all interdependent. This has the consequence that clear limits on $T_c$ are set, thus excluding high temperature superconductivity within the BCS approach. Bednorz and Müller [3], instead, observed that some oxide superconductors showed a rather high value of $T_c$ in spite of an extremely low carrier density. This led them to conclude, that an unconventionally large coupling between the carriers and the lattice had to be present which is beyond the framework considered in BCS theory. As a possible route for achieving such a strong coupling, they assumed that polaron formation could be at work, where specifically the concept of Jahn-Teller polarons [12] was considered [13–15]. A polaron is a quasi-particle where the carrier and the lattice deformation cloud, as depicted in Fig. 5, form a new entity which can travel through the lattice. In order to achieve superconductivity two of these quasi-particles combine to a bipolaron which in contrast to the conventional Cooper pair (Fig. 6 (left)) is correlated over a few nm only. This limits the number of pairs within a unit volume and overlap between them is unlikely (Fig. 6 (right)).

The formation of this quasi-particle and its bound state is possible only when there is an extremely strong interaction supporting a high superconducting transition temperature. Simultaneously, the density of the carriers can be rather small. The crucial issue of these considerations was the search for a suitable material. While Bednorz and Müller first started with the perovskite $SrTiO_3$ they soon found out that it is indeed possible to induce superconductivity in it, however with a disappointingly small transition temperature (<3 K) [16]. The next compound was again a perovskite type oxide, a nickelate, where superconductivity could not be verified. Finally, they came along with the cuprates, the high-temperature superconductors. Their starting concept of the bipolaron mechanism of high-temperature superconductivity was thus realized, remained, however, under strong debate and a variety of novel and exotic theories followed rapidly after this discovery, where no consensus exists until today. It has to be emphasized that in spite of the existing controversies in this field, the Nobel laureates initiated a number of experiments to verify their concept, mainly by concentrating on isotope experiments. While in conventional superconductors the isotope effect on $T_c$ is inversely proportional to the square root of the ionic mass and constant, in cuprates it depends on the number of carriers and eventually even vanishes [17, 18]. The latter fact has been advocated to require novel theories, however, also polaron based theories are able to explain it [13–15]. Besides of the isotope effect on $T_c$ a number of novel isotope effects were observed, as e.g. one on the penetration depth and the energy gap.

## 5 Development of technological applications

Since the discovery of superconductiviy in 1911 by Heike Kammerlingh Onnes and Gilles Holst [1] tremendous advances have been made both in understanding the nature of superconductivity and the materials which show it. A second revolution in the field of superconductivity happened in 1986 with the finding of the cuprate high-temperature superconductors [3] with transition temperatures well above the boiling point of liquid nitrogen (77 K or –196°C). In the cover story of *Time Magazine* (11 May 1987) the importance of this discovery was stated as a "*starting breakthrough that could change our world*" with euphoric predictions of all kinds of technical applications [19, 20]. In fact, it was already obvious from the very beginning in 1911 that superconductivity bares a great potential for a variety of unique technologies, including superconducting magnets, transmission power cables, electrical motors and generators, energy storage devices, maglev trains, and various kinds of superconducting sensors, to give only a few examples [2, 19–24]. Figure 10 shows a schematic overview of the broad spectrum of already existing applications using high-temperature superconductors (HTSs).

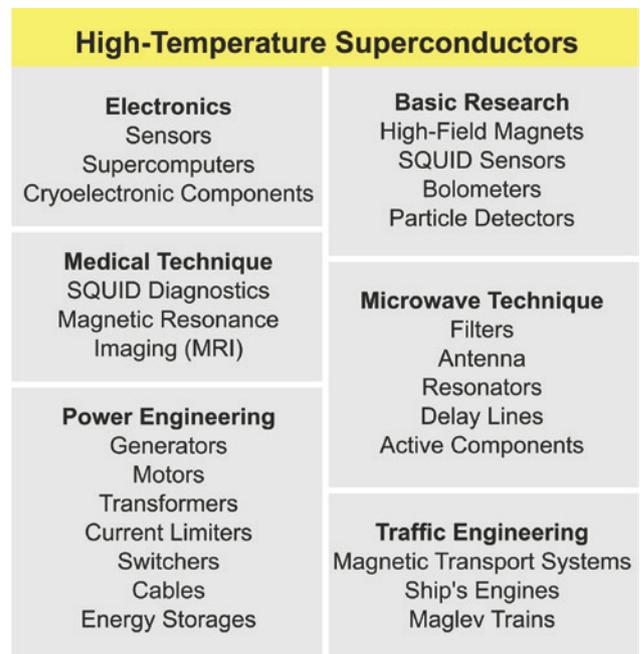

**Fig. 10:** Schematic overview of possible applications of cuprate high-temperature superconductors (HTSs).

After the discovery of superconductivity it took quite some time to find superconducting materials which are suitable for power applications. Such a material must have a rather high $T_c$ and carry a high electrical current (high critical current density $j_c$) in a high magnetic field (high critical magnetic field $H_c$) (see Table 3). In 1954 the intermetallic compound $Nb_3Sn$ was found to be such a material, followed by other A15 compounds and NbTi (up to now still the most used material for technical applications) [2, 21, 22]. Using these materials it was possible to build superconducting magnets which produce the high magnetic fields needed e.g. in magnetic resonance imaging (MRI) scanners for medical diagnostics (see Fig. 11).

The discovery of the cuprate HTSs [3] opened a new era of applications. It became in principle possible to operate superconducting devices at much higher working temperatures than conventional ones, i.e. above the boiling point of liquid nitrogen (77 K or –196°C) which is much cheaper and much easier to obtain and to handle than liquid helium (4 K or –269°C). However, the high transition temperature $T_c$ is not the only crucial parameter for superconducting wires useful for electrical power engineering. The wires must carry a high current density (high citical curent density $j_c$) and withstand a high magnetic field (high critical magnetic field $H_c$). Cuprate HTSs are so-called type II superconductors which are characterized by two critical magnetic fields: the lower critical magnetic field $H_{c_1}$ and the upper critical magnetic field $H_{c_2}$. For technical applications $H_{c_2}$ is relevant. This is illustrated in Fig. 12 where the critical parameters $T_c$, $j_c$, and $H_{c_2}$ of a low-temperature superconductor (LTS) are compared to those of a cuprate HTS. Note that in the ideal case all parameters of a HTS are higher than those of a LTS which makes the former so interesting for technical applications. However, there are some other drawbacks for the application of cuprate HTSs as discussed below.

**Table 3:** Crucial parameters of a superconducting material relevant for technical applications.

| |
| --- |
| High criticall temperature $T_c$ (>77 K) |
| High critical current density $j_c$ |
| High critical magnetic field $H_c$ |
| Good and stable mechanical properties |
| Simple and low-cost production |

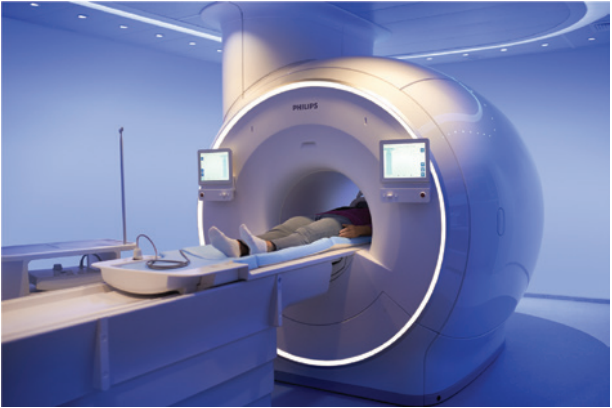

**Fig. 11:** MRI scanner for medical applications (Philips Ingenia 3.0T MR System). The wires of the superconducting magnet (3 Tesla) consist of the conventional low-temperature superconductor NbTi. (Courtesy of *Philips*).

Besides the cuprate HTSs other superconducting materials were discovered after 1986 which are also suitable for applications, such as magnesium diboride $MgB_2$ with $T_c = 39$ K (–234°C) [9] and a series of iron-based compounds called pnictides with a $T_c$ up to 56 K (–221°C) [10].

In the following we will focus on applications using cuprate HTSs. Their essential novelty is clearly the high superconducting transition temperature $T_c$ compared to those of LTSs. This implies a drastic reduction of cooling costs, a major factor for large scale technologies. The main criteria a superconducting material should fulfill for applications are summarized in Table 3. One essential hurdle to handle with was to produce useful HTS materials in the form of long robust and flexible cables for high power engineering as well as stable high-quality thin films for low power electronics and sensor devices. HTSs are very complex materials consisting of several elements with multiple phases. They are very brittle ceramics which makes it difficult to optimize all the crucial material aspects. Furthermore, cuprate HTSs are layered superconductors with a short coherence length $\xi$. The supercurrents predominantly flow in the copper-oxygen planes. Consequently, the critical parameters $j_c$ and $H_c$ as well as the coherence length $\xi$ are highly anisotropic. To achieve a sufficiently high current density for high power application, the grains in the wires have to be well aligned over distances as long as several 100 m. This is a difficult material science problem, but nowadays such HTS cables are available and are tested in real electric power grids (see Figs. 13 and 14).

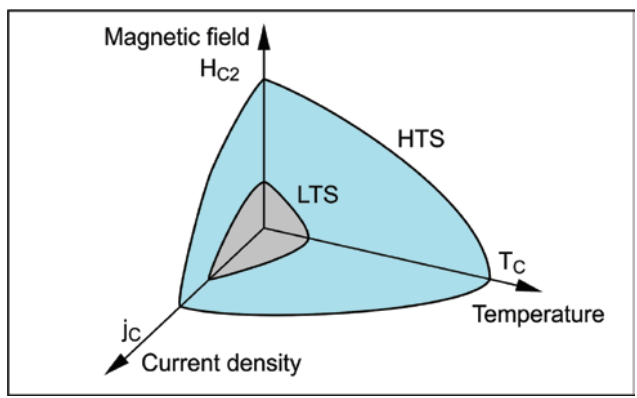

**Fig. 12:** Schematic diagram of the critical parameters (critical temperature $T_c$, critical current density $j_c$, and upper critical magnetic field $H_{c_2}$) of a superconductor (LTS, low-temperature superconductor; HTS, high-temperature superconductor).

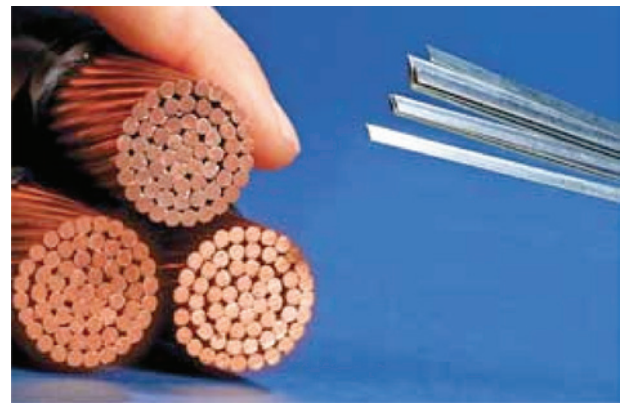

**Fig. 13:** Superconducting cable made of a HTS material (right) is able to carry 150 times the electrical current of traditional copper wire of the same size (left). (Courtesy of *American Superconductor Corp.*).

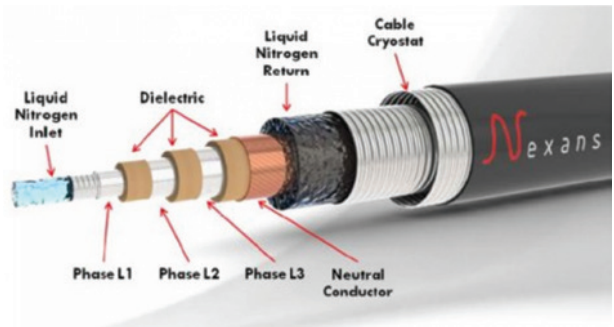

**Fig. 14:** HTS power transmission cable developed by *Nexans SuperConductors GmbH*. The concentric arrangement of three phases L1, L2, and L3 allows a very compact cable design. The conducting parts of the cable (L1, L2, L3) consists of a HTS material. The cable is cooled by liquid nirogen.

Since 1986 tremendous progress has been made to fabricate useful HTS materials for various kinds of applications. A schematic overview of HTS applications is shown in Fig. 10 and detailed descriptions of HTS applications are given in Refs. [19, 20, 22–24]. Here we only present a small selection of some unique applications of cuprate HTSs.

Certainly the greatest commercial potential for HTS materials is in electric power applications which require long and flexible high-quality HTS wires and cables with the desired physical properties. Examples of such power applications include high-power transmission cables, high-field superconducting magnets, motors and generators, synchronous condensers, transformers, and fault-current limiters [19, 20, 22–24]. Such wires and cables are developed by several companies around the world and are available on the market. Figure 14 shows an example of a HTS power transmission cable consisting of three cores for the three phases of electric power.

In 2014 the world's longest HTS cable was integrated to the power grid of Essen (Germany) and put into real operation. The system consists of a one-kilometer 10 kV (2300 A) HTS cable (Fig. 14), including a joint and a fault-current limiter, cooled by liquid nitrogen, and connecting two substations in the city center of Essen [25]. It was designed to replace a high-voltage cable (110 kV) with a medium-voltage (10 kV) cable, thus saving space in substations by eliminating the need for transformers and other bulky equipment. The particularly efficient and space-saving technology transports five times more electricity than conventional cables with almost no losses. The field test of this novel power grid system may be pathfinding for planning future energy supply systems of cities. This new system also contains a fault-current limiter based on HTS technology which must be able to withstand fault currents. The simplest form of it makes use of the fact that above a certain critical current, a superconductor becomes a normal resistive conductor, and consequently the high current through the system can be switched to low. Various types of fault-current limiters were developed by several companies and are available on the market. Another rather advanced field of applications are rotating electromechanical machines such as motors, generators, and synchronous condensers. All these machines contain copper coils which partly can be replaced by coils made of HTS material, especially the rotor coils. HTS rotating machinery has several advantages compared to conventional designs. The losses can be reduced by a factor of two including the cooling of the HTS system. The higher magnetic fields produced by HTS rotor coils allow a very compact design of a motor concerning weight and size. For instance, *American Superconductor* has designed a 36.5 MW HTS ship propulsion engine that has only 20% of the weight and volume of a conventional one.

HTSs are also promising materials to realize efficient energy storage devices. There are in principle two types of such devices [20, 24]: (i) A superconducting magnetic energy storage (SMES) device consists basically of a superconducting coil in which electric energy is stored in the form of a frictionless flowing electric supercurrent in the coil. The use of a HTS material for the coil allows a compact and efficient construction of the device. (ii) In a flywheel energy storage (FES) device a rotor (flywheel) is accelerated to a high speed and the energy in the system is conserved as rotational energy. When energy is extracted (added) from the system, the rotational speed of the flywheel is reduced (increased) according to the law of conservation of energy. The flywheels spinning at high speeds (20 000 to over 50 000 rpm) are suspended by frictionless magnetic bearings based on HTS technology.

An interesting application of HTS technology is employed in superconducting magnetic levitation (SCMaglev) trains developed by the *Central Japan Railway Company (JR Central)* and the company's *Railway Technical Research Institute*. The SCMaglev railway system is based on the principle of magnetic repulsion between the track and the cars. The coils on the car that produce the magnetic field are made of HTS material. In 2015 a SCMaglev train reached a speed of 603 km/h on a 42.8 km magnetic-levitation test track in Japan (https://www.youtube.com/watch?v#equal#dQ1eljTcmK4). This is the world record for manned passenger trains. Commercial SCMaglev service is scheduled for 2027 between Tokyo and Nagoya being 286 km apart. The SCMaglev train would run at a top speed of 500 km/h and connect the two cities in 40 min, less than half the present travel time in a Shinkansen.

There are a number of applications which do not require HTS materials in long wire forms, high power levels or high magnetic fields. These are mainly related to the field of electronic devices and often call for small thin HTS films.

Due to the increasing demand for wireless communication, the largest commercial use of HTS electronic devices is as filters in cell phone base stations where the low microwave resistivity and noise of HTS thin films operating at 77 K (liquid nitrogen) enables a broader signal range than conventional metallic filters. HTS materials are also used for high-Q resonant cavities in high-energy particle accelerators.

Another important application of HTS electronics is the SQUID (Superconducting QUantum Interference Device) which makes use of the Josephson effect (tunneling of Cooper pairs through a narrow barrier between two superconductors, called a Josephson junction). SQUIDs are very effective sensors to detect tiny magnetic fields and play an important role in various commercial applications such as in basic research, technical material testing and characterization, and medical diagnostics. As an example, SQUIDS can detect with good spatial resolution the magnetic fields generated by human heart currents or currents in the brain.

Predictions for the future are always uncertain, but it may be stated that superconductivity very likely will continue to provide unexpected discoveries and surprises, ranging from new types of superconducting materials to novel kinds of technologies. However, unless there is no other way to solve a technological problem, devices based on superconductors are always in strong competition with technological solutions which are more practical and cheaper.

# 6 Outlook, challenges and a little science fiction

## 6.1 Theories advocating room-temperature superconductivity

Since the discovery of high-temperature superconductivity in cuprates, the hope to achieve room-temperature (RT) superconductivity has considerably increased together with massive research activities in this direction. As has been mentioned before, the high pressure study of $H_2S$ aimed in this direction, however these results have the drawback, that the enormous pressures used there are unfavorable for useful applications. In addition, the reported values of $T_c$ of 200 K and even higher, have not been reproduced until today. This means that the chase for $T_c$ values higher than 135 K continues and that novel approaches beyond BCS theory or more exotic theories are needed. Also, the till now considered inorganic materials must be abandoned and new directions pursued.

More than 60 years ago, in a pioneering work, Fröhlich [6] predicted that low-dimensional systems exhibiting incommensurate charge density waves could be candidates for high-temperature superconductivity. Even though superconductivity was rapidly verified in these compounds, the maximum value of $T_c < 10$ K could not be exceeded.

Shortly before the BCS theory Matthias [26], who had discovered most of the then existing superconductors, established an empirical rule for further searches, namely based on the average filling of the transition metal d-bands. This rule had more exceptions than applicable results, however yielded unexpected surprises in the field. In this context, also the layered graphitic superconductor compounds were discovered, and for the first time it was pointed out that their two-dimensional structure could be the major ingredient to make them superconduct [27].

Two years after the publication of the BCS theory, Suhl et al. [28] introduced extensions of the model by taking into account that the Fermi surface of more realistic compounds may have more than one band being relevant for superconductivity. They proposed a modification of the theory in terms of two-band superconductivity. The intriguing feature of it lies in the fact that the superconducting transition temperature is always higher as compared to the single-band approach and may even reach values exceeding 200 K in the weak coupling limit. While being introduced early on, an experimental realization of it followed only in the 1980[th] in Nb doped $SrTiO_3$ [16]. While this case has long been considered as an exception, meanwhile a variety of two-band superconductors, with the most prominent case of $MgB_2$ [9], have been discovered. Especially, for $MgB_2$ it has been clearly demonstrated that the single band approach yields substantially lower values of $T_c$ than experimentally observed [29].

An innovative proposal was made by Little [30] in 1965, when he suggested that one-dimensional structures would be favorable for room temperature or even higher temperature superconductivity. In particular, as a specific example he proposed organic polymers in which the pair binding is not provided by the ionic nuclei vibrations, but instead by electronic vibrations. In such a case the inverse electronic mass, being much smaller than the ionic mass, sets the scale for $T_c$ and admits for very high transition temperatures.

Around the same time Ginzburg developed a theoretical frame using the BCS theory to achieve RT superconductivity [31]. He concentrated on the specific ingredients of the BCS theory, namely, the electron-phonon interaction, the density of states, the Coulomb repulsion, and the cutoff energy.

Simultaneously, Ashcroft considered possibilities to enhance $T_c$ within the framework of the BCS theory [32]. Since the phonon energy is an important ingredient here, he proposed to increase this by using light elements with high vibration frequencies and specifically hydrogen. Since hydrogen is difficult to metallize even at high pressure, his predictions remained unverified until today, with the exception of $H_2S$ mentioned before [11].

Another decade later and 10 years before the discovery of cuprate superconductivity, Ginzburg and Kirzhnits [33] edited a book entitled *High-Temperature Superconductivity* where a variety of materials including the "Little" organic ones, were discussed as candidates for room temperature superconductivity. Shortly after this in Japan intense research efforts were bundled in a new project focusing on the same issue. A further attempt to find high temperature superconductivity was initiated by T. Geballe and collaborators in organizing the first international workshop on "*Mechanism and Materials of Superconductivity*".

While results of studies of oxide superconductors remained elusive for many years and only few were reported in comparison to more than 1000 other superconductors, the observation of superconductivity in ternary oxides with rather high values of $T_c$ gave rise to enter new pathways in the research of high-$T_c$ superconductivity including the discovery of cuprates. All ternary or more complex superconducting oxides exhibit an astonishingly small density of states as compared to normal metals or other superconductors. Thus their elevated values of $T_c$ must stem from an extremely strong pairing glue which might be beyond the BCS phonon mechanism as suggested by Bednorz and Müller and inferring a bipolaronic scenario [34]. However, the lack of direct evidence for this specific mechanism also gave rise to the upcoming of exotic and novel pairing models, where specifically the resonating valence bond model, first introduced by Pauling [35] in 1948, has received intense attention [36]. In addition, the magnetic ground states of cuprates and pnictides was taken as evidence that magnetic fluctuations give rise to high-temperature superconductivity. In this case the attractive interaction is not limited by the phonon frequency spectrum but can have much higher energies and correspondingly yield much higher $T_c$'s.

## 6.2 Reports of room-temperature superconductivity: USOs

Unidentified superconducting objects (USOs) with transition temperatures around room temperature (RT) have been reported throughout the 1970s and 1980s and more frequently after the discovery of the cuprates. A certain boom arises nowadays with reports of transition temperatures far above RT initiated by the high pressure studies on $H_2S$ [11]. An early example is the Al-C-Al sandwich structure [37], where the strong dependence of the current on a small external magnetic field resembles the Josephson DC tunneling current. This phenomenon was observed at 300 K and interpreted as superconductivity, being unconfirmed until today. However, the report initialized further work in the field where Anilin Black (AB) was postulated as a candidate for RT superconductivity ($T_c = 295.7$ K) and related to the above sandwich structure, since highly conductive quasi-one-dimensional domains are formed [38]. The idea of Little was taken up by Ladik and Bierman [39] to postulate high-temperature superconductivity through virtual excitations of the electrons in one chain of double stranded DNA. Related to these ideas and reports are investigations of superconductivity in small metallic crystallites which are embedded in vapor-deposited thin films. Upon coating these crystallites with various dielectrics, a large surface volume ratio is achieved where the electrons in the vicinity of the surface have different properties as compared to the ones from the bulk [40]. Specifically, it can excite polarization waves inside the dielectric which have much higher frequencies than the Debye frequency. Consequently, surface superconductivity is expected with $T_c \simeq 10^2 – 10^4$ K. These considerations have been taken up more recently by Eagles [41], who suggested probable RT superconductivity in small systems. The examples include $CdF_2$-based sandwich structures embedded in boron doped layers. For such a system zero resistance was reported at 319 K, large diamagnetism at even higher temperature and tunneling evidence for an energy gap was given [42]. Related compounds are multi-walled carbon nanotubes or mats of single-walled or multi-walled nanotubes, where $T_c$ values of 700 K were observed from resistivity and diamagnetic susceptibility measurements [43]. Another example is given by narrow channels through films of oxidized atactic polypropylene (OAPP) with superconductivity persisting up to 429 K [44].

As mentioned above, the ideas of Ashcroft have been taken up many times over the years and have seemingly been realized in $H_2S$ under extremely high pressure. Latest new results in favor of this proposal have been presented by Drozdov et al. in lanthanum hydrides with a record $T_c$ of 215 K again under high pressure [45]. Similar findings have been made by the group of Hemley in $LaH_{10}$ with onset temperatures even around room temperature [46]. In both preprints, the resistivity drop is observed, however, diamagnetic susceptibility data are missing.

A novel type of superconducting phase has been reported by Kawashima where several forms of graphite and single-layer graphene wetted with alkanes exhibit superconductor-like properties above room temperature under ambient pressure [47]. The novelty is here that superconductivity is restricted to a certain temperature range below which it disappears. In a very recent paper [48] the observation of superconductivity at ambient temperature and pressure conditions in films and pellets of a nanostructured material that is composed of silver particles embedded into a gold matrix has been reported. Specifically, upon cooling below 236 K at ambient pressures, the resistance of sample films drops below $10^{-4}$ Ω. Below the transition temperature, samples become strongly diamagnetic, with volume susceptibilities as low as −0.056. Also recipes are discussed how to tune the transition to temperatures higher than room temperature. The above examples may belong to the category of USOs, since they have not been reproduced till now or susceptibility data are absent. Also reduced $WO_3$ has to be mentioned in this context, where superconductivity was reported at much lower temperature, $T_c \approx 120$ K [49], but could never be confirmed by other groups.

## 6.3 Novel application tools for the future

As pointed out above superconductivity – and in particular high-temperature superconductivity – admits for diverse applications ranging from energy transport and harvesting to medical diagnostics (see Fig. 10). Until to date, however, all known superconductors need to be cooled either by liquid nitrogen or by liquid helium, where the latter is expensive. For $H_2S$ ultrahigh pressures are required. The challenges in the field of superconductivity are to reach room temperature (RT) values of $T_c$ and to arrive at a material which is easily employable. The fact that there is no unique consensus on the pairing mechanism of high-temperature superconductivity limits this search, since no predictive routes can be used in the search for new materials. Besides this drawback, it is also desirable to look for materials consisting of abundant elements in order to decrease the manufacturing costs. Further requisites are easy manufacturing processes, stability, and flexibility. To meet all these criteria it might be anticipated that the research in this field needs novel approaches and new mindsets beyond our todays knowledge.

However, nature is full of surprises and from the present scientific knowledge of superconductivity it is not impossible that superconductivity at room temperature or above might be realizable 1 day. If this were achieved the implications for applications of room-temperature (RT) superconductors are enormous and high ranking with respect to energy harvesting and production costs as long as ambient pressure can be used. Imagine RT superconductors with the desired properties listed in Table 3 would be available, this would change not only our technological world, but also our daily life completely. Superconductivity would be present in most of our technical devices and amazing novel types of technologies would be possible. Several *Gedanken* experiments would be possible which might be realized with RT superconductors in the future:

An electrical current that flows forever without loosing any energy means transport of electric power with virtually no losses in the transmission cables. A high-power grid could contain high-power components based on RT superconductivity technology such as transformers, dynamic synchronous condensers, and fault-current limiters. In addition, a superconducting wire carrying a current that never diminishes would act as a perfect superconducting magnetic energy storage (SMES) device operating at RT. Unlike conventional batteries, which degrade over time, in a RT SMES device energy can be captured and stored for ever without any appreciable losses. The same would apply for a flywheel energy storage (FES) device containing bearings made of a RT superconductor material. Furthermore, superconducting magnets with incredibly high magnetic fields can be realized for scientific (e.g. high-field laboratory magnets), technical (e.g. nuclear fusion reactors to produce electric power) and medical applications (high-resolution MRI scanners) with relatively low production costs. RT superconductivity would also have a great impact on rotating electromechanical machines such as motors and generators which may be constructed in a very compact and light way, since no cooling system is required. This is especially important for a novel generation of electric cars where the engine as well as the electric storage device may be based on RT superconductor technology. A revolution would also happen in the development of superconducting magnetic levitation (SCMaglev) trains as a key transport system of the future. Further, it would also have an immense impact on low-power applications of superconductivity, such as sensors, SQUIDs, detectors, and filters used in various fields of technology ranging from medical applications to information technology. In addition, novel approaches to realize very efficient and compact Josephson and quantum computers would be possible.

Although RT superconductivity is still a dream, many scientists believe that 1 day it will be realized. This would unleash astonishing novel technologies and alter our daily life tremendously.

**Acknowledgements:** We gratefully acknowledge encouraging and constructive discussions with Prof. K. A. Müller. Special thanks are devoted to Prof. A. Simon for fruitful discussions and continuous support of our work in the field of high-temperature superconductivity and related research topics. We also kindly thank R. Noack for preparing the figures.

# Graphical synopsis

Annette Bussmann-Holder and Hugo Keller
**High-temperature superconductors: underlying physics and applications**

https://doi.org/10.1515/znb-2019-0103
Z. Naturforsch. 2019; x(x)b: xxx–xxx

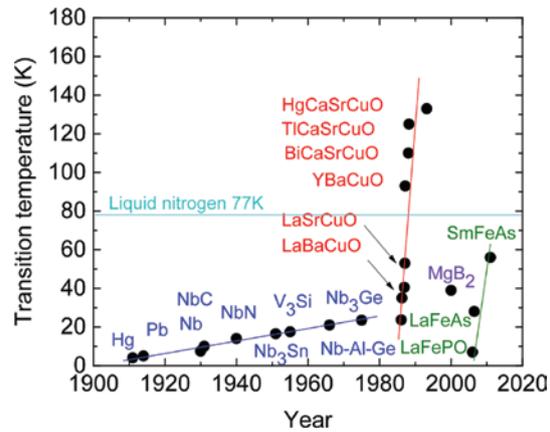